\documentclass[aps,prb,twocolumn,amsmath,amssymb,nofootinbib,superscriptaddress,floatfix]{revtex4-2}
\usepackage[dvips]{graphicx}
\usepackage{latexsym}
\usepackage{amsmath}
\usepackage{amsfonts}
\usepackage{amssymb}
\usepackage{bm}
\usepackage{color}
\usepackage{txfonts}
\usepackage{float}
\usepackage{url}
\usepackage[colorlinks=true, urlcolor=blue, linkcolor=blue, citecolor=blue, pdftex]{hyperref}
\usepackage{ulem}
\begin{document}
	\newcommand{\fig}[2]{\includegraphics[width=#1]{#2}}
	\newcommand{\la}{{\langle}}
	\newcommand{\ra}{{\rangle}}
	\newcommand{\dg}{{\dagger}}
	\newcommand{\upa}{{\uparrow}}
	\newcommand{\dna}{{\downarrow}}
	\newcommand{\ab}{{\alpha\beta}}
	\newcommand{\ias}{{i\alpha\sigma}}
	\newcommand{\ibs}{{i\beta\sigma}}
	\newcommand{\hH}{\hat{H}}
	\newcommand{\hn}{\hat{n}}
	\newcommand{\hc}{{\hat{\chi}}}
	\newcommand{\hU}{{\hat{U}}}
	\newcommand{\hV}{{\hat{V}}}
	\newcommand{\br}{{\bf r}}
	\newcommand{\bk}{{{\bf k}}}
	\newcommand{\bq}{{{\bf q}}}
	\def\gsim{~\rlap{$>$}{\lower 1.0ex\hbox{$\sim$}}}
	\setlength{\unitlength}{1mm}
	\newcommand{{\vhf}}{$\chi^\text{v}_f$}
	\newcommand{{\vhd}}{$\chi^\text{v}_d$}
	\newcommand{{\vpd}}{$\Delta^\text{v}_d$}
	\newcommand{{\ved}}{$\epsilon^\text{v}_d$}
	\newcommand{{\vved}}{$\varepsilon^\text{v}_d$}
	\newcommand{{\tr}}{{\rm tr}}
	\newcommand{\pprl}{Phys. Rev. Lett. \ }
	\newcommand{\pprb}{Phys. Rev. {B}}

\title {Kramers Fulde-Ferrell state and superconducting spin diode effect}
\author{Yi Zhang}
\email{zhangyi821@shu.edu.cn}
\affiliation{Department of Physics, Shanghai University, Shanghai 200444, China}

\author{Ziqiang Wang}
\email{ziqiang.wang@bc.edu}
\affiliation{Department of Physics, Boston College, Chestnut Hill, MA 02467, USA}

\date{\today}

\begin{abstract}
We study a novel equal-spin pairing state with opposite center of mass momentum for each spin polarization.
This state, dubbed a Kramers Fulde-Ferrell (KFF) state, respects time-reversal symmetry and can be realized in a one-dimensional system with spin-orbit coupling and nearest neighbor attraction.
We find that the KFF state supports
nonreciprocal spin transport for both bulk superconductor and Josephson junctions.
In addition to the spin Josephson diode effect,
the charge transport
is controlled by intriguing dynamics of bound states whose transitions can be manipulated by the length of the KFF superconductor. The KFF state is relevant for embedded quantum structures in monolayer Fe-based superconductors and dissipationless superconducting spintronics.
\end{abstract}
\maketitle

\section{Introduction}
Recent experimental observations of the diode effect in superconductors~\cite{ando,sde_NbSe2,kim_2021,tTLG} and Josephson junctions (JJ)~\cite{ali,Bau_2022,Baumgartner_2022,NiTe2,tbg} have stimulated the research of nonreciprocal transport properties in superconducting (SC) systems. Following the proposal of SC diode effect \cite{hu},
the so-called $\phi_0$ Josephson state has been extensively studied~\cite{buzdin_2008,yuli_2013,yuli_2014,julia_2015,kou_2016,Marco_2019,phi0_2013,phi0_2016,phi0_2017,phi0_2018,phi0_2020,phi0_2021,phi0_2022} as a possible mechanism to realize nonreciprocal transport in JJs.
More recently, many theoretical proposals~\cite{nagaosa1,hejun,yuan,yanse,fu_2022,tTLGt,yi2022,ruben_2022} have been put forward for SC diode in bulk superconductors.
In particular, the finite-momentum pairing Flude-Ferrell-Larkin-Ovchinnikov state~\cite{FF,LO} is believed to provide a physical mechanism, since the order parameter of Fulde-Ferrell (FF) state $\Delta(\textbf{r})=\Delta e^{i\textbf{q}\cdot\textbf{r}}$ can directly generate a difference in the critical current along and against the direction of $\textbf{q}$, leading to SC diode effect \cite{fu_2022,yanse}.
The FF order, also known as helical superconductivity, can be realized in noncentrosymmetric superconductors with spin-orbit coupling (SOC)
and time reversal symmetry breaking fields~\cite{Smidman_2017,gorkov_2001,victor_2002,AGTERBERG200313,dim_2003,kaur_2005,kaur_2007,dim_2007,sam_2008,yanase_2008,mich_2012,sek_2013,houzet_2015,yuan_2019}.

So far, the study of SC diode effect
focused on the nonreciprocity of charge transport.
One may wonder if there exists a similar nonreciprocal property in spin transport in certain SC systems.
From the symmetry point of view, SC diode effect in charge transport, where critical currents in opposite directions have different magnitudes, requires the system to break both inversion and time-reversal (${\cal T}$) symmetry, since the charge current operator changes sign under either inversion or ${\cal T}$.
While the spin current operator changes sign under inversion, it is invariant under ${\cal T}$.
Thus, the nonreciprocity in spin transport only requires breaking inversion symmetry and can be realized in ${\cal T}$ invariant superconductors.

In this article, we propose a novel SC state that can realize nonreciprocal spin transport. This state has
equal-spin pairing and a FF type of order parameter $\Delta_{\sigma}(\textbf{r})=\Delta e^{i\sigma\textbf{Q}\cdot\textbf{r}}$, with opposite Cooper pair center of mass momentum for opposite spin polarizations as shown schematically in Fig.~\ref{fig:pd}(a). We term this SC state as a Kramers FF (KFF) state since ${\cal T}$ symmetry is maintained.
Such FF state has pairing field with only one $\textbf{Q}$ vector in each pairing channel, which is translational invariant unlike the Larkin-Ovchinnikov state also known as the pair density wave state, where the pairing field has both $\textbf{Q}$ and $-\textbf{Q}$ vectors in each channel so that the pairing order parameters varies in space.
We demonstrate that the KFF state can be realized in a meanfield theory of a concrete model describing a spin-orbit coupled chain with nearest neighbor attractions as illustrated in Fig.~\ref{fig:pd}(a,b). The nonzero $\textbf{Q}$ pairing across the Fermi points in Fig.~\ref{fig:pd}(a) is enabled by the SOC split bands.
We study the condition to realize the nonreciprocal spin transport where the critical {\em spin} current along positive and negative directions are unequal in magnitude
for both bulk SC state and Josephson junction structures. Moreover, we find intriguing properties and rich phases in the charge transport across JJs of the KFF state, which can be realized by simply changing the length of the SC chain.
Similar pairing state was also studied in the two dimensional honeycomb system where the valley degree of freedom plays the role of spin here~\cite{lee_2019}.

This article is organized as follows. We start with the introduction of our model Hamiltonian and meanfield formulation for the KFF state in Sec.~\ref{sec:model}. In Sec.~\ref{sec:bulk}, we discuss the nonreciprocal spin transport for the bulk SC with KFF order, which is followed by the discussion of the hidden inversion symmetry that is related to the nonreciprocal spin transport in Sec.~\ref{sec:hidden} Then we study the transport properties of the Josephson junction structure constructed from the KFF state in Sec.~\ref{sec:JJ} and discuss various phases realized in the charge transport across JJs of the KFF state in Sec.~\ref{sec:jje}. We finalize the discussion in Sec.~\ref{sec:dis}.

\section{Formulation}
\label{sec:model}
\subsection{Model Hamiltonian}
We first consider a one-dimensional (1D) spin-orbit coupled chain with nearest neighbor attraction described by the Hamiltonian
\begin{equation}
	\hat{H}=-\sum_{i,j,\sigma}t_{ij}c_{i\sigma}^{\dagger}c_{j\sigma}+\alpha\sum_{i}i\sigma c_{i\sigma}^{\dagger}c_{i+1\sigma}+h.c.-V\sum_{i}n_{i}n_{i+1}
	\label{eq:htb}
\end{equation}
where $t_{ij}$ are hopping parameters
up to the 2nd neighbor ($t_1$ and $t_2$) and $\alpha$ describes a nearest neighbor SOC. In 1D, SOC leaves a conserved spin quantum number which is taken to be the spin quantization axis along the chain direction.
The nearest neighbor attraction $V$ responsible for SC order can be decomposed into equal-spin and opposite-spin pairing channels as 
\begin{equation}
	\begin{split}
		H_{I}&=-V\sum_{i}n_{i}n_{i+1}=-V\sum_{i,\sigma\sigma^{\prime}}c_{i\sigma}^{\dagger}c_{i+1\sigma^{\prime}}^{\dagger}c_{i+1\sigma^{\prime}}c_{i\sigma}
		\\
		&=-\frac{V}{N_{c}}\sum_{k,k^{\prime},q,\sigma\sigma^{\prime}}e^{i(k-k^{\prime})}c_{k+\frac{q}{2}\sigma}^{\dagger}c_{-k+\frac{q}{2}\sigma^{\prime}}^{\dagger}c_{-k^{\prime}+\frac{q}{2}\sigma^{\prime}}c_{k^{\prime}+\frac{q}{2}\sigma}
		\\
		&=-\frac{V}{N_{c}}\sum_{k,k^{\prime},q,\sigma}\sin k\sin k^{\prime}c_{k+\frac{q}{2}\sigma}^{\dagger}c_{-k+\frac{q}{2}\sigma}^{\dagger}c_{-k^{\prime}+\frac{q}{2}\sigma}c_{k^{\prime}+\frac{q}{2}\sigma}
		\\
		&-\frac{V}{N_{c}}\sum_{k,k^{\prime},q,\sigma}e^{i(k-k^{\prime})}c_{k+\frac{q}{2}\sigma}^{\dagger}c_{-k+\frac{q}{2}\bar{\sigma}}^{\dagger}c_{-k^{\prime}+\frac{q}{2}\bar{\sigma}}c_{k^{\prime}+\frac{q}{2}\sigma}
	\end{split}
	\label{eq:Hamil}
\end{equation}
where the first term corresponds to the attraction between the electrons with the same spin and the second term corresponds to the attraction between the electrons with opposite spins and $N_c$ is the number of sites. If we further define the two pairing operators in equal-spin and opposite-spin channels as
\begin{equation}
	\begin{cases}
		\hat{\Delta}_{\parallel,q,\sigma}=\frac{1}{N_{c}}\sum_{k} i\sin k c_{-k+\frac{q}{2}\sigma}c_{k+\frac{q}{2}\sigma}\\
		\hat{\Delta}_{\perp,q,\sigma}=\frac{1}{N_{c}}\sum_{k}e^{-ik} c_{-k+\frac{q}{2}\bar{\sigma}} c_{k+\frac{q}{2}\sigma}
	\end{cases}
	\label{eq:ansatz}
\end{equation}
Eq.~\ref{eq:Hamil} can be written as
\begin{equation}
	H_{I}=-N_{c}V_{1}\sum_{q}\hat{\Delta}_{\parallel,q,\sigma}^{\dagger}\hat{\Delta}_{\parallel,q,\sigma}-N_{c}V_{2}\sum_{q}\hat{\Delta}_{\perp,q,\sigma}^{\dagger}\hat{\Delta}_{\perp,q,\sigma}
\end{equation}
where these two terms correspond to the pairing channels with equal and opposite spin respectively and here we denote the attraction in these two channels as $V_{1}$ and $V_{2}$. 
While $V_1=V_2=V$ in the original model in Eq.~(\ref{eq:htb}), we consider here a more general model where the effective attraction $V_1$ and $V_2$
can be different. The equal-spin pairing can be induced in embedded quantum structures in high-T$_c$ superconductors due to spatial symmetry breaking~\cite{yi_2021}, such as along the atomic line defects in monolayer FeTeSe ~\cite{chen_2020}.
Then the total Hamiltonian becomes
\begin{equation}
	\hat{H}=\sum_{k,\sigma} \varepsilon_{k,\sigma} c_{k\sigma}^{\dagger}c_{k\sigma}-N_{c}V_{1}\sum_{q}\hat{\Delta}_{\parallel,q,\sigma}^{\dagger}\hat{\Delta}_{\parallel,q,\sigma}-N_{c}V_{2}\sum_{q}\hat{\Delta}_{\perp,q,\sigma}^{\dagger}\hat{\Delta}_{\perp,q,\sigma}
	\label{eq:htbq}
\end{equation}
where
\begin{equation}
	\varepsilon_{k\sigma}= -2t_{\alpha}\cos(k-\sigma\theta_{\alpha}) - 2t_{2}\cos(2k)
\end{equation}
is the band dispersion with
\begin{equation}
	\begin{cases}
		t_{\alpha}=\sqrt{t_{1}^{2}+\alpha^{2}}
		\\
	    \theta_{\alpha}=\arctan(\alpha/t_{1})
	\end{cases}
\end{equation}
which determines the positions of the Fermi points.

\subsection{Meanfield decoupling}
\begin{figure}
	\begin{center}
		\fig{3.4in}{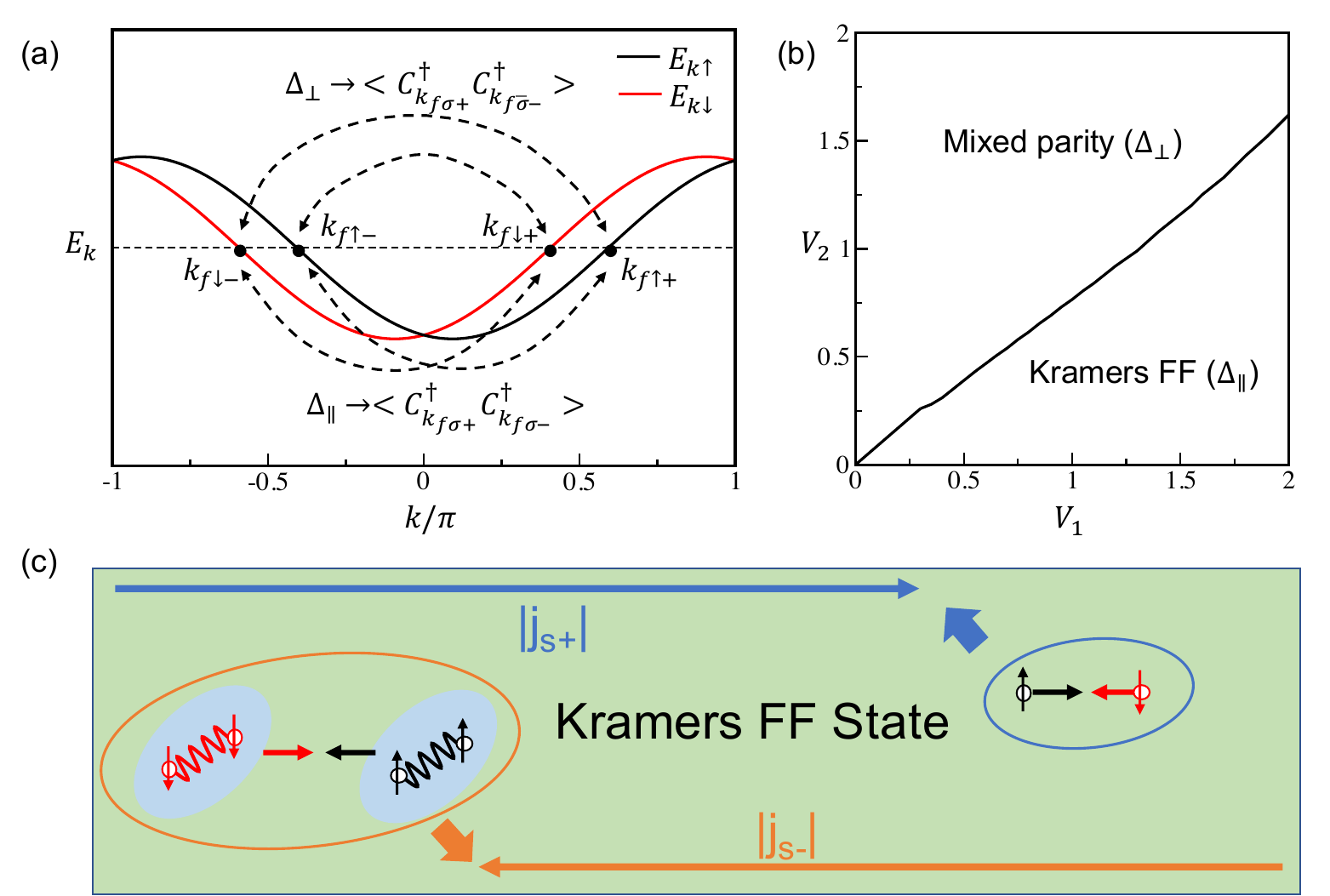}\caption{(a) Schematic band structure of the spin-orbit coupled 1D chain, showing the two pairing channels across the Fermi points. (b) Phase diagram obtained from meanfield calculations with parameters $t_{1}=1$, $t_{2}=-0.5$, $\alpha=0.4$, $\mu=-0.53$. (c) Schematic illustration of nonreciprocal spin transport due to different critical spin currents $\vert j_{s,c-}\vert > \vert j_{s,c+}\vert$ in ``$+$''and ``$-$'' directions. A spin current $ \vert j_{s,c+} \vert  <\vert j_{s,\pm}\vert  < \vert j_{s,c-}\vert $ flows as a dissipationless supercurrent in ``$-$'' direction (marked by $\vert j_{s-}\vert$), but can only be transported as a normal dissipative current in ``$+$'' direction (marked as $\vert j_{s+}\vert$).
\label{fig:pd}}
	\end{center}
	\vskip-0.5cm
\end{figure}

From the structure of Fermi points shown in Fig.~\ref{fig:pd}(a), we can solve the model in Eq.~(\ref{eq:htbq}) within a meanfield approximation assuming the following meanfield ansatz
\begin{equation}
	\begin{cases}
		\left\langle \hat{\Delta}_{\parallel,Q,\uparrow}\right\rangle =\left\langle \hat{\Delta}_{\parallel,-Q,\downarrow}\right\rangle =\Delta_{\parallel}\\
		\left\langle \hat{\Delta}_{\perp,0,\uparrow}\right\rangle =\Delta_{\perp}e^{i\phi_{\perp}}\\
		\left\langle \hat{\Delta}_{\perp,0,\downarrow}\right\rangle =-\Delta_{\perp}e^{-i\phi_{\perp}}
	\end{cases}
\end{equation}
Here, in equal-spin pairing channel, electrons with up (down) spin pair into the FF state with a nonzero center of mass momentum Q(-Q). The resulting Kramers doublet ensures that ${\cal T}$ symmetry is preserved. In opposite-spin pairing channel, electrons with up and down spins form zero-momentum pairs, which is in general a mixture of $s$-wave and $p_{z}$-wave pairing depending on the phase $\phi_{\perp}$. Specifically, $\phi_{\perp}=0$ corresponds to $s$-wave pairing and $\phi_{\perp}=\frac{\pi}{2}$ corresponds to $p_{z}$-wave pairing, while other values give rise to a mixed parity state.

After the meanfield decoupling, the meanfield Hamiltonian can be written as
	\begin{equation}
		\begin{split}
			\hat{H}_{MF}-\mu\hat{N}&=\sum_{k\sigma}\left(\varepsilon_{k\sigma}-\mu\right)c_{k\sigma}^{\dagger}c_{k\sigma}
			\\
			&-V_{1}\Delta_{\parallel}\sum_{k\sigma}(-i\sin k)c_{k+\frac{\sigma Q}{2}\sigma}^{\dagger}c_{-k+\frac{\sigma Q}{2}\sigma}^{\dagger}
			\\
			&-2V_{2}\Delta_{\perp}\sum_{k}c_{k\uparrow}^{\dagger}c_{-k\downarrow}^{\dagger}\cos\left(k+\phi_{\perp}\right)+h.c.
			\\
			&+2N_{c}(V_{1}\Delta_{\parallel}^{2}+V_{2}\Delta_{\perp}^{2})
		\end{split}
		\label{eq:hmft}
	\end{equation}
In the Nambu basis $\psi_{k}^{\dagger}=\left(c_{k+\frac{Q}{2}\uparrow}^{\dagger},c_{k-\frac{Q}{2}\downarrow}^{\dagger},c_{-k+\frac{Q}{2}\uparrow},c_{-k-\frac{Q}{2}\downarrow}\right)$, Eq.~\ref{eq:hmft} can be written as
	\begin{equation}
		H_{MF}-\mu N=\frac{1}{2}\sum_{k}\psi_{k}^{\dagger}h_{k}\psi_{k}+2N_{c}V_{1}\Delta_{\parallel}^{2}+2N_{c}V_{2}\Delta_{\perp}^{2}-\mu N_{c}
	\end{equation}
	with
\begin{widetext}
	\begin{equation}
		h_{k}=\left[\begin{array}{cccc}
			\varepsilon_{k+\frac{Q}{2},\uparrow}-\mu & 0 & 2iV_{1}\Delta_{\parallel}\sin k & -2V_{2}\Delta_{\perp}\cos(k+\frac{Q}{2}+\phi_{\perp})\\
			0 & \varepsilon_{k-\frac{Q}{2},\downarrow}-\mu & 2V_{2}\Delta_{\perp}\cos(-k+\frac{Q}{2}+\phi_{\perp}) & 2iV_{1}\Delta_{\parallel}\sin k\\
			-2iV_{1}\Delta_{\parallel}\sin k & 2V_{2}\Delta_{\perp}\cos(-k+\frac{Q}{2}+\phi_{\perp}) & -\varepsilon_{-k+\frac{Q}{2},\uparrow}+\mu & 0\\
			-2V_{2}\Delta_{\perp}\cos(k+\frac{Q}{2}+\phi_{\perp}) & -2iV_{1}\Delta_{\parallel}\sin k & 0 & -\varepsilon_{k-\frac{Q}{2},\downarrow}+\mu
		\end{array}\right]
	\end{equation}
\end{widetext}
This meanfield Hamiltonian can be solved self-consistently for a fixed chemical potential $\mu$ and various values of $Q$ and $\phi_{\perp}$ with the self-consistent equations
\begin{equation}
	\begin{cases}
		\Delta_{\parallel}=\frac{1}{2N_{c}}\sum_{k\sigma}i\sin k\left\langle c_{-k+\frac{\sigma Q}{2},\sigma}c_{k+\frac{\sigma Q}{2},\sigma}\right\rangle \\
		\begin{split}
			\Delta_{\perp}&=\frac{1}{N_{c}}\sum_{k}\cos(k+\phi_{\perp})\left\langle c_{-k\downarrow}c_{k\uparrow}\right\rangle
			\\ &=\frac{1}{N_{c}}\sum_{k}\cos(k+\frac{Q}{2}+\phi_{\perp})\left\langle c_{-k-\frac{Q}{2}\downarrow}c_{k+\frac{Q}{2}\uparrow}\right\rangle
		\end{split} 
	\end{cases}
\end{equation}
and the ground state is determined by the states with the lowest free energy density $\Omega=\frac{1}{N_{c}}\left\langle H_{MF}-\mu N\right\rangle$  which also determines the value of $Q$ and $\phi_{\perp}$. 

We perform the calculation with a general set of parameter and the obtained meanfield phase diagram in Fig.~\ref{fig:pd}(b) shows that the novel KFF state is a more stable ground state than the mixed parity state when  $V_1$ is larger than $V_2$.
We thus focus on the KFF state driven by equal-spin pairing and investigate its many intriguing properties.
A more detailed analysis of the meanfield phase diagram as well as the behavior of the order parameters are shown in Appendix.~\ref{appendixA}. The detailed calculation for the mixed parity state is also shown in Appendix.~\ref{appendixE}.

\section{Nonreciprocal spin transport and Spin diode effect in bulk KFF state}
\label{sec:bulk}
The meanfield Hamiltonian in KFF state becomes
\begin{equation}
	\begin{split}
		\hat{H}_{MF}-\mu\hat{N}&=\sum_{k\sigma}\left(\varepsilon_{k\sigma}-\mu\right)c_{k\sigma}^{\dagger}c_{k\sigma}
		\\
		&+V_{1}\Delta_{\parallel}\sum_{k\sigma}i\sin kc_{k+\frac{\sigma Q}{2},\sigma}^{\dagger}c_{-k+\frac{\sigma Q}{2},\sigma}^{\dagger}+h.c.+2N_{c}V_{1}\Delta_{\parallel}^{2}
	\end{split}
	\label{eq:mf}
\end{equation}
which can be written in the Nambu basis $\psi_{k\sigma}^{\dagger}=\left(c_{k+\frac{\sigma Q}{2},\sigma}^{\dagger},c_{-k+\frac{\sigma Q}{2},\sigma}\right)$ as
\begin{equation} \hat{H}_{MF}-\mu\hat{N}=\frac{1}{2}\sum_{k\sigma}\psi_{k\sigma}^{\dagger}h_{k,Q,\sigma}\psi_{k\sigma}+2N_{c}V_{1}\Delta_{\parallel}^{2}-\mu N_{c}
	\label{eq:hkmf}
\end{equation}
where,
\begin{equation}
	h_{k,Q,\sigma}=\left[\begin{array}{cc}
		\varepsilon_{k+\frac{\sigma Q}{2},\sigma}-\mu & 2iV_{1}\Delta_{\parallel}\sin k\\
		-2iV_{1}\Delta_{\parallel}\sin k & -\varepsilon_{-k+\frac{\sigma Q}{2},\sigma}+\mu
	\end{array}\right]
\end{equation}
is block diagonal in spin space, leading to the eigenenergy as
\begin{equation}
	\begin{split}
		E_{k\sigma,\pm}&=\frac{1}{2}\left(\varepsilon_{k+\frac{\sigma Q}{2},\sigma}-\varepsilon_{-k+\frac{\sigma Q}{2},\sigma}\right)
		\\
		&\pm\sqrt{\left[\frac{1}{2}(\varepsilon_{k+\frac{\sigma Q}{2},\sigma}+\varepsilon_{-k+\frac{\sigma Q}{2},\sigma})-\mu\right]^{2}+4V_{1}^{2}\Delta_{\parallel}^{2}\sin^{2}k}
	\end{split}
	\label{eq:ekpm}
\end{equation}
and we have $E_{k\sigma,\pm}=E_{-k\bar{\sigma},\pm}$ due to the ${\cal T}$ symmetry. Then the free energy density at zero temperature $\Omega(\Delta_{\parallel},Q)$ can be calculated as
\begin{equation}
	\begin{split}
		\Omega(\Delta_{\parallel},Q)&=\frac{1}{N_{c}}\left\langle \hat{H}_{MF}-\mu\hat{N}\right\rangle
		\\ &=\frac{1}{2N_{s}}\sum_{k\sigma,n=\pm}E_{k\sigma,n}\Theta(-E_{k\sigma,n})+2V_{1}\Delta_{\parallel}^{2}-\mu	
	\end{split}
    \label{eq:omegak}
\end{equation}
where  $\Theta(x)$ the Heaviside step function.
The order parameter $\Delta_{\parallel}$ for a given $Q$ can be determined self-consistently by minimizing $\Omega(\Delta_{\parallel},Q)$ with respect to $\Delta_{\parallel}$, leading to the self-consistency equation
\begin{equation}
	\Delta_{\parallel}=\frac{1}{2N_{c}}\sum_{k\sigma}i\sin k\left\langle c_{-k+\frac{\sigma Q}{2},\sigma}c_{k+\frac{\sigma Q}{2},\sigma}\right\rangle
\end{equation}
The optimal $Q$ value can be further determined by minimizing $\Omega(\Delta_{\parallel},Q)$ with respect to $Q$, i.e.,
$\partial_{Q}\Omega(\Delta_{\parallel},Q)=0$.
Because the latter is directly related to the spin current carried by the KFF state (see Appendix.~\ref{appendixB}) $j_{s}(Q)=\sum_{\sigma}\frac{\sigma}{2}j_{\sigma}=\partial_{Q}
\Omega(\Delta_{\parallel},Q,T)$,
the ground state with optimized
$Q=Q_{0}$ does not carry any net spin current since
$\partial_{Q}\Omega(\Delta_{\parallel},Q_{0},T)=0$. When
the KFF state is driven out of equilibrium into a state with
$Q\neq Q_0$, a nonzero applied spin current $j_{s}(Q\neq Q_0)=j_{s}\neq0$ is realized.
Throughout the remaining text, we define the charge and spin currents in units of $\frac{e}{\hbar}$ and unity, respectively.
Since the spin current carrying state has equal but opposite center of mass momentum $\pm Q$ for Cooper pairs in opposite spin channels, the charge current always vanishes due to $\cal T$ symmetry.
The {\em critical} spin currents in $+$ and $-$ directions are determined by the maximum and minimum values of $j_{s}(Q)$ sustained by the SC state according to
$j_{s,c+}=\max_{Q}[j_{s}(Q)]$ and $j_{s,c-}=\min_{Q}[j_{s}(Q)]$. When $\left|j_{s,c+}\right|\neq\left|j_{s,c-}\right|$, the SC state enables nonreciprocal spin transport as shown in Fig.~\ref{fig:pd}(c).

We performed meanfield calculations using two sets of parameters. In the first case, we set $t_2=0$ and obtain analytically that the ground state has $Q_{0}=2\theta_{\alpha}=\sigma(k_{f\sigma,+}+k_{f\sigma,-})$, consistent with the SOC split bands
where electrons pair across the Fermi points $k_{f\sigma,\pm}$ in the same spin sector, giving rise to finite Cooper pair momenta $\sigma Q_{0}$ with more details shown in Appendix.~\ref{appendixC}.
The zero temperature free energy in Eq.~(\ref{eq:omegak}) and the current density from its momentum derivative are calculated numerically and
plotted in Figs.~\ref{fig:je}(a, c) as a function of $Q$.
The critical spin currents $j_{s,c\pm}$ are determined by the maximum and minimum values of the spin current density.
Fig.~\ref{fig:je}(c) shows that the two critical momenta $Q_{\pm}$ at which the spin current reaches critical values coincide with the two momenta where the SC order parameter $\Delta_{\parallel}(Q_{\pm})$ vanishes. In this case with $t_{2}=0$, we find that the critical spin currents $j_{s,c+}=-j_{s,c-}$, as shown in Fig.~\ref{fig:je}(a), and the spin transport is reciprocal. The absence of nonreciprocal spin transport turns out to be due to a hidden inversion symmetry when $t_2=0$.

\section{Hidden inversion symmetry}
\label{sec:hidden}
To demonstrate the hidden inversion symmetry, we can perform a local gauge transformation $c_{i\sigma}^{\dagger}\rightarrow e^{-\frac{i}{2}\sigma Qx_{i}}d_{i\sigma}^{\dagger}$, corresponding to $c_{k\sigma}^{\dagger}\rightarrow d_{k-\frac{\sigma Q}{2},\sigma}^{\dagger}$ in momentum space.
Then the meanfield Hamiltonian Eq.~\ref{eq:mf} becomes
	\begin{equation}
		\begin{split}
			\hat{H}_{MF}-\mu\hat{N}&=\sum_{k\sigma}\left(\varepsilon_{k\sigma}-\mu\right)d_{k-\frac{\sigma Q}{2},\sigma}^{\dagger}d_{k-\frac{\sigma Q}{2},\sigma}
			\\
			&+V_{1}\Delta_{\parallel}\sum_{k\sigma}i\sin kd_{k,\sigma}^{\dagger}d_{-k,\sigma}^{\dagger}+h.c.+2N_{c}V_{1}\Delta_{\parallel}^{2}
			\\
			&=\sum_{k\sigma}\left(\varepsilon_{k+\frac{\sigma Q}{2},\sigma}-\mu\right)d_{k,\sigma}^{\dagger}d_{k,\sigma}
			\\
			&+V_{1}\Delta_{\parallel}\sum_{k\sigma}i\sin kd_{k,\sigma}^{\dagger}d_{-k,\sigma}^{\dagger}
			+h.c.+2N_{c}V_{1}\Delta_{\parallel}^{2}
		\end{split}
	\end{equation}
Since in the absence of $t_{2}$, $\varepsilon_{k+\frac{\sigma Q}{2},\sigma}=-2t_{\alpha}\cos(k+\frac{\sigma Q}{2}-\sigma\theta_{\alpha})$, we can see that in the new basis, the only inversion breaking term $\theta_{\alpha}$ owing to the spin-orbit coupling $\alpha$, is cancelled if $Q=Q_{0}=2\theta_{\alpha}$. In other words, when $Q=Q_{0}=2\theta_{\alpha}$, the inversion symmetry can be recovered in the new basis, where the meanfield Hamiltonian becomes
\begin{equation}
	\begin{split}
		\hat{H}_{MF}-\mu\hat{N}&=\sum_{k\sigma}\left(-2t_{\alpha}\cos k-\mu\right)d_{k,\sigma}^{\dagger}d_{k,\sigma}
		\\
		&+V_{1}\Delta_{\parallel}\sum_{k\sigma}i\sin kd_{k,\sigma}^{\dagger}d_{-k,\sigma}^{\dagger}+h.c.+2N_{c}V_{1}\Delta_{\parallel}^{2}
	\end{split}
\end{equation}
The transformed Hamiltonian describes two spin-degenerate $p$-wave Kitaev chains~\cite{Kitaev_2001} with inversion symmetry in each spin sector. It is precisely this hidden inversion symmetry that forbids the nonreciprocal property of spin current, since this hidden inversion symmetry changes the sign of the current operator $\hat{j}_{\sigma}$, while keeping the total Hamiltonian invariant, which guarantees a one to one correspondence between the positive and negative current, such that $j_{s,c+}$ and $j_{s,c-}$ have to have the same magnitude. 

\begin{figure}
	\begin{center}
		\fig{3.4in}{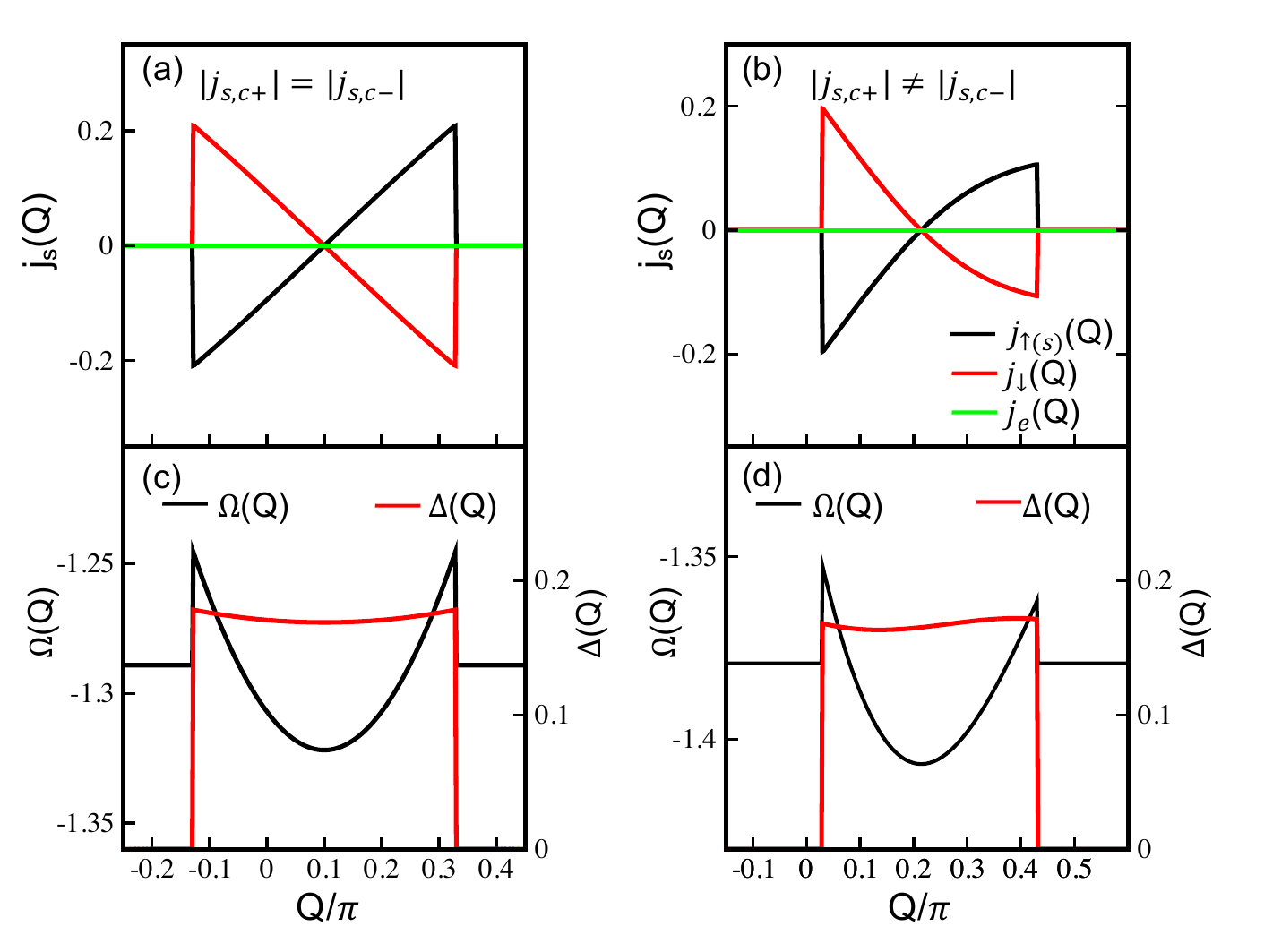}\caption{Spin current density (a, b), free energy and SC order parameter (c, d) as a function of momentum $Q$.
Parameters for (a) and (c): $t_{2}=0$ and $t_{1}=1$, $\alpha=\tan(\frac{\pi}{20})$, $\mu=0$, $V_1=2$. Parameters for  (b) and (d): $t_{2}=-0.5$ and $t_{1}=1$, $\alpha=0.4$, $\mu=-0.53$, $V_1=2$.
The spin current is defined as $j_{s}=\frac{1}{2}(j_{\uparrow}-j_{\downarrow})$.
			\label{fig:je}}
	\end{center}
	\vskip-0.5cm
\end{figure}

In the case with finite $t_2$, the dispersion of the noninteracting Hamiltonian reads $\varepsilon_{k\sigma}=-2t_{\alpha}\cos(k-\sigma\theta_{\alpha})-2t_{2}\cos(2k)$, and then the Fermi momentum $k_{f\sigma,\pm}$ no longer have a closed form and the sum of the two Fermi momentum belonging to the same spin polarization is incommensurate in general. Here, we can immediately see that the gauge transformation above can not recover the inversion symmetry as above, since now $\varepsilon_{k+\frac{\sigma Q}{2},\sigma}=-2t_{\alpha}\cos(k+\frac{\sigma Q}{2}-\sigma\theta_{\alpha})-2t_{2}\cos(2k+\sigma Q)$, and the inversion breaking phase of the two cosine function $\frac{\sigma Q}{2}-\sigma\theta_{\alpha}$ and $\sigma Q$ cannot be cancelled by $Q$ simultaneously, so that there is no hidden inversion symmetry that forbids the presence of the nonreciprocal spin transport.
Therefore, the transformed model describes the two p-wave Kitaev chains with complex hoppings, which are time-reversal counterparts of each other but not identical and can be written in the real space as
\begin{equation}
	\begin{split}
		\hat{H}_{MF}-\mu\hat{N}&=
		\sum_{i\sigma}\Bigl(-t_{\alpha}e^{i\sigma(\frac{Q}{2}-\theta_{\alpha})}d_{i,\sigma}^{\dagger}d_{i+1,\sigma}
		-t_{2}e^{i\sigma Q}d_{i,\sigma}^{\dagger}d_{i+2,\sigma} 
		\\
		&+V_{1}\Delta_{\parallel}d_{i,\sigma}^{\dagger}d_{i+1,\sigma}^{\dagger}+h.c.
		-\mu d_{i,\sigma}^{\dagger}d_{i,\sigma} \Bigl)+2N_{c}V_{1}\Delta_{\parallel}^{2}
	\end{split}
\end{equation}
Indeed, the results shown in Fig.~\ref{fig:je}(b) confirms that critical spin currents are nonreciprocal with $j_{s,c+}=0.11$ and $j_{s,c-}=-0.20$ along $+$ and $-$ directions, respectively.
As a result, a spin current $j_{s}$
satisfying $0.11<|j_{s}|<0.20$ flows as dissipationless supercurrent in the negative direction since $|j_{s}|<|j_{s,c-}|$, but can only be transported as a dissipative normal current in the positive direction since $|j_{s}|>|j_{s,c+}|$. This SC spin diode effect is shown schematically in Fig.~\ref{fig:pd}(c).

\section{Josephson junctions}
\label{sec:JJ}
We next study the transport properties of Josephson chains consisting of a normal metal sandwiched between two KFF superconductors depicted in Fig.~\ref{fig:jj}(a) with more detailed setups shown in Appendix.~\ref{appendixD}.
We consider the general case with nonzero $t_2$ in the KFF state.

\subsection{Spin Diode effect with spin-independent phase bias}
\begin{figure}
	\begin{center}
		\fig{3.4in}{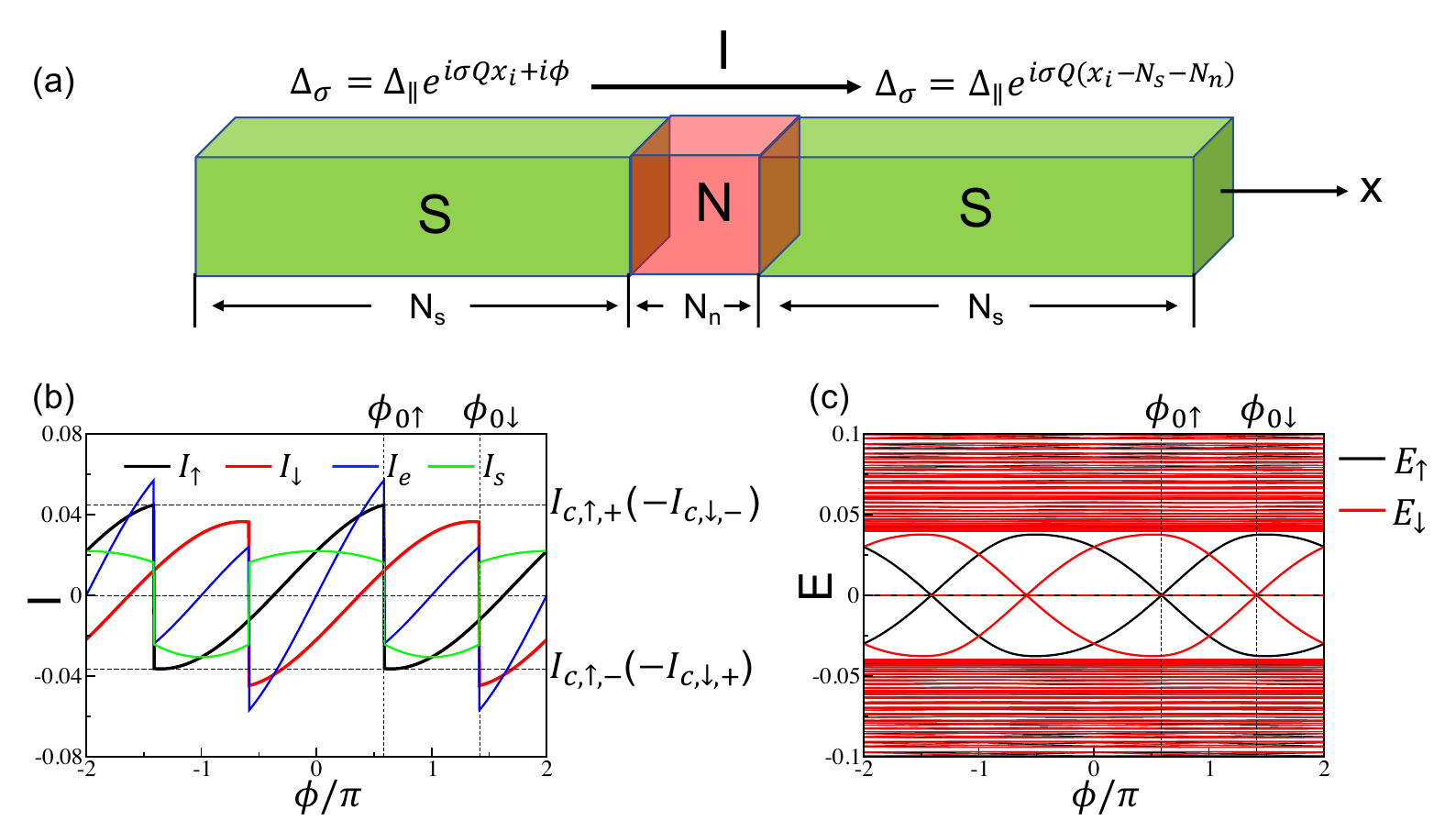}\caption{(a) Schematics of an S-N-S Josephson junction.
			The lengths of the KFF-SC (S) and normal metal (N) regions are $N_{s}$ and $N_{n}$. (b) Current phase relation for the Josephson chain with a typical KFF-SC order determined self-consistently for parameters: $t_{1}=1$, $t_{2}=-0.5$, $\alpha=0.5$, $\mu=-0.7$, $V_{1}=1$, leading to $\Delta_{\parallel}=0.063$ and $Q=0.372\pi$. Upper and lower horizontal dashed lines indicate critical current $I_{\uparrow,c+}$ and $I_{\uparrow,c-}$ along ``$\pm$'' directions.
			(c) Energy-phase spectrum of the Josephson chain in (b). Two vertical dashed lines in (b) and (c) correspond to phase bias $\phi_{0\uparrow}$ and $\phi_{0\downarrow}$ where the bound states cross zero, leading to jumps in the Josephson current shown in (b). $N_{s}=319$ and $N_{n}=3$. The ``N'' region has nearest neighbor hopping $t_N=1$. The couplings between ``N'' and ``S'' regions are described by $t_L=t_R=1$ as defined in Appendix.~\ref{appendixD}.
			\label{fig:jj}}
	\end{center}
	\vskip-0.5cm
\end{figure}

We first study the transport properties of JJ with
{\em spin-independent} phase bias $\phi$.
The Josephson currents can be calculated by the formula
\begin{equation}
	I(\phi)=\frac{2e}{\hbar}\frac{\partial\Omega_{Jc}(\phi)}{\partial\phi}
\end{equation}
where $\Omega_{Jc}(\phi)$ is the free energy of the system~\cite{beenakker}.
The results for both spin components
and total charge and spin currents are shown in Fig.~\ref{fig:jj}(b). Due to ${\cal T}$ symmetry, the spin-dependent critical currents satisfy $I_{\sigma,c+}=-I_{\bar{\sigma},c-}$ such that charge current is reciprocal $\vert I_{e,c+}\vert = \vert I_{e,c-}\vert$.
In contrast, the critical current for each spin is asymmetric, i.e.
$|I_{\sigma,c+}|\neq|I_{\sigma,c-}|$, giving rise to nonreciprocal Josephson spin currents $\vert I_{s,c+}\vert \neq \vert I_{s,c-}\vert$  and the SC spin diode effect.

\subsection{Spin diode effect in the Josephson junction with spin phase}
\label{sec:sd}
\begin{figure}[h]
	\begin{center}
		\fig{3.2in}{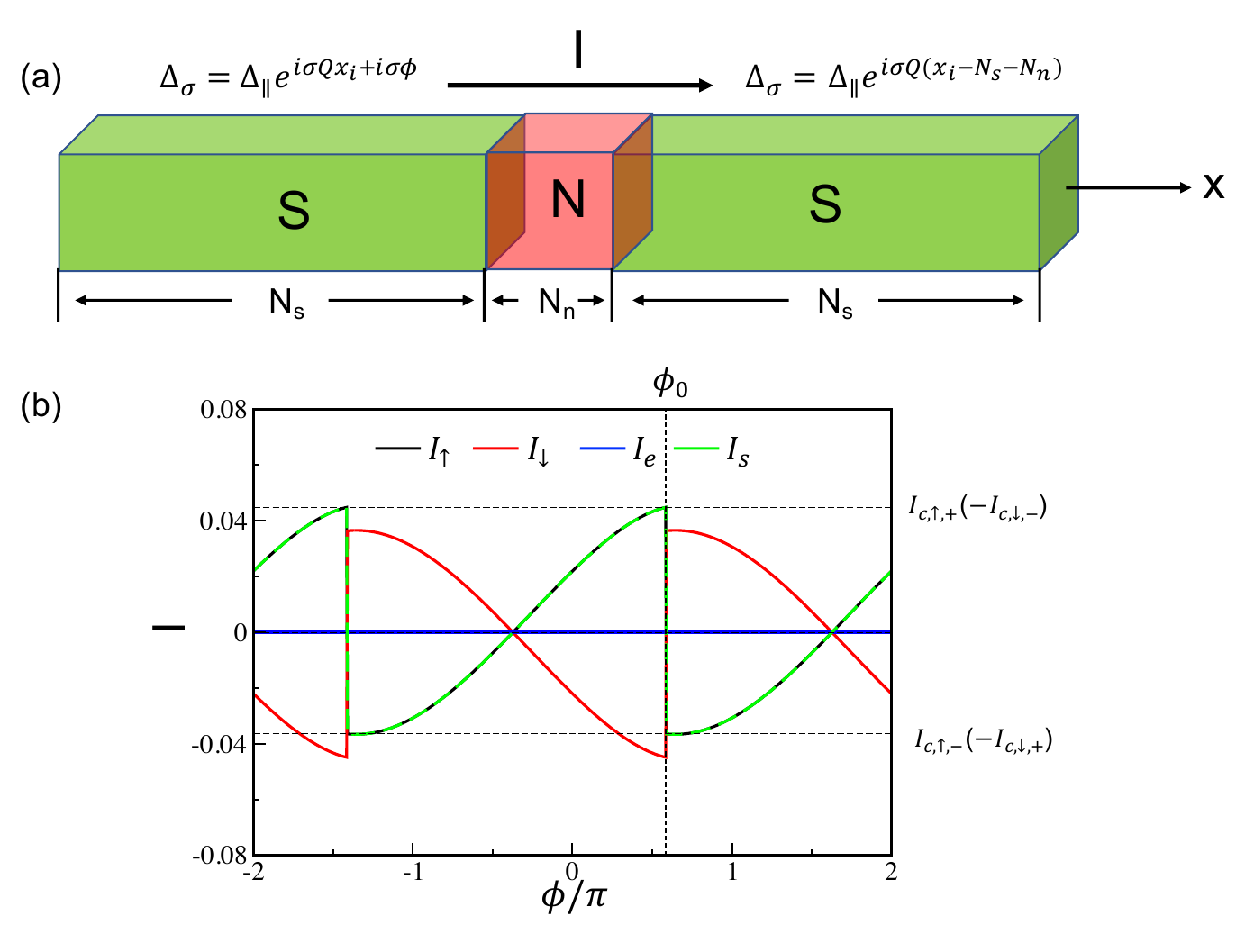}\caption{(a) Schematics of a S-N-S Josephson junction with a spin phase in $z$ direction.
			The lengths of the KFF-SC (S) and normal metal (N) regions are $N_{s}$ and $N_{n}$. (b) Current phase relation for the Josephson chain with a typical KFF-SC order determined self-consistently for parameters: $t_{1}=1$, $t_{2}=-0.5$, $\alpha=0.5$, $\mu=-0.7$, $V_{1}=1$, leading to $\Delta_{\parallel}=0.063$ and $Q=0.372\pi$. Upper and lower horizontal dashed lines indicate critical current $I_{\uparrow,c+}$ and $I_{\uparrow,c-}$ along ``$\pm$'' directions.
			$N_{s}=319$ and $N_{n}=3$ are used in the calculations, and the ``N'' region has nearest neighbor hopping $t_N=1$. The coupling between the ``N'' and ``S'' regions are described by $t_L=t_R=1$ as defined in Appendix.~\ref{appendixD}.
			\label{fig:jjs}}
	\end{center}
	\vskip-0.5cm
\end{figure}

The spin diode effect with nonreciprocal spin current by applying a spin-dependent phase bias known as the spin phase~\cite{sun_2022} is obvious.
The spin phase was first introduced in Ref.~\onlinecite{sun_2022}. In the current case, we consider the spin phase in the $z$ direction, which corresponds to the spin-dependent phase bias $\phi_{\sigma}=\sigma\phi$ applied to the Josephson junction. We consider the same parameter set as that shown in Fig.~\ref{fig:jj} and the resulting current phase relation is shown in Fig.~\ref{fig:jjs}. Here, since the spin phase $\phi_{\sigma}=\sigma\phi$ still respects the $\cal{T}$ symmetry, the current for the opposite spin polarization are always equal in magnitude and opposite in the direction, i.e., $I_{\uparrow}(\phi)=-I_{\downarrow}(\phi)$, so that the total charge current $I_e(\phi)$ always vanishes and the total spin current $I_s(\phi)=\frac{1}{2}(I_{\uparrow}(\phi)-I_{\downarrow}(\phi))=I_{\uparrow}(\phi)$, which is nonreciprocal as long as the hidden inversion symmetry is broken by finite $t_2$. The nonreciprocal spin current of the Josephson junction with spin phase is inherited from the nonreciprocal spin transport of the bulk KFF state. 

\section{Length controlled phase transitions in charge transport}
\label{sec:jje}
An intriguing feature
in Fig.~\ref{fig:jj}(b) is the phase difference between Josephson currents in two spin sectors, which is clearly revealed in the energy spectrum plotted in Fig.~\ref{fig:jj}(c).
Apart from the continuum states outside the SC gap, there are eight in-gap states. Among them, four are at exactly zero energy, corresponding to two pairs of Majorana zero modes (one for each spin) located at the two ends of the chain due to $p$-wave nature of the KFF state~\cite{Kitaev_2001}, which do not contribute to Josephson current. The other four in-gap states are Andreev bound states of the S-N-S junction.
For junctions made of conventional superconductors, the energies of the bound states cross at $\phi=\pi$~\cite{rmp_jj1}, corresponding to zero modes trapped by $\pi$-junction~\cite{sauls_2018}. In the current KFF JJs, the energy spectrum of the bound states for two spin species shifts in opposite directions as shown in Fig.~\ref{fig:jj}(c). As a result, the phase bias where the bound states cross zero shifts from $\pm\pi$ to $\pm\phi_{0\sigma}$, where the Josephson current jumps due to branch switching as shown in Fig.~\ref{fig:jj}(b).

Such phase differences have a great impact on charge transport.  The charge current $I_{e}=I_{\uparrow}+I_{\downarrow}$ crosses zero at both $\phi=0$ and $\phi=\pi$ with positive slopes, indicating the free energy $\Omega_{Jc}(\phi)$ of the junction reaches a local minimum at both $\phi=0$ and $\pi$.
This state is called a $\mathbf{0}^{\prime}$ or $\pmb{\pi}^{\prime}$ state depending on the momentum of the global minimum.
It was studied previously in JJs where two superconductors are coupled through an Anderson impurity~\cite{rozh_1999} or a magnetic quantum dot~\cite{siano_2004}.
Here, these remarkable states are realized in ${\cal T}$-invariant systems due to the novel KFF SC order. Remarkably, we find that distinct Josephson junction states ($\mathbf{0}$, $\pmb{\pi}$, $\mathbf{0}^{\prime}$ and $\pmb{\pi}^{\prime}$) 
can all be realized by tuning the phase shift,
which can be easily achieve by changing the length of the SC region. The definition of these states are listed in Table.~\ref{table1}.

\begin{figure*}
	\begin{center}
		\fig{7.0in}{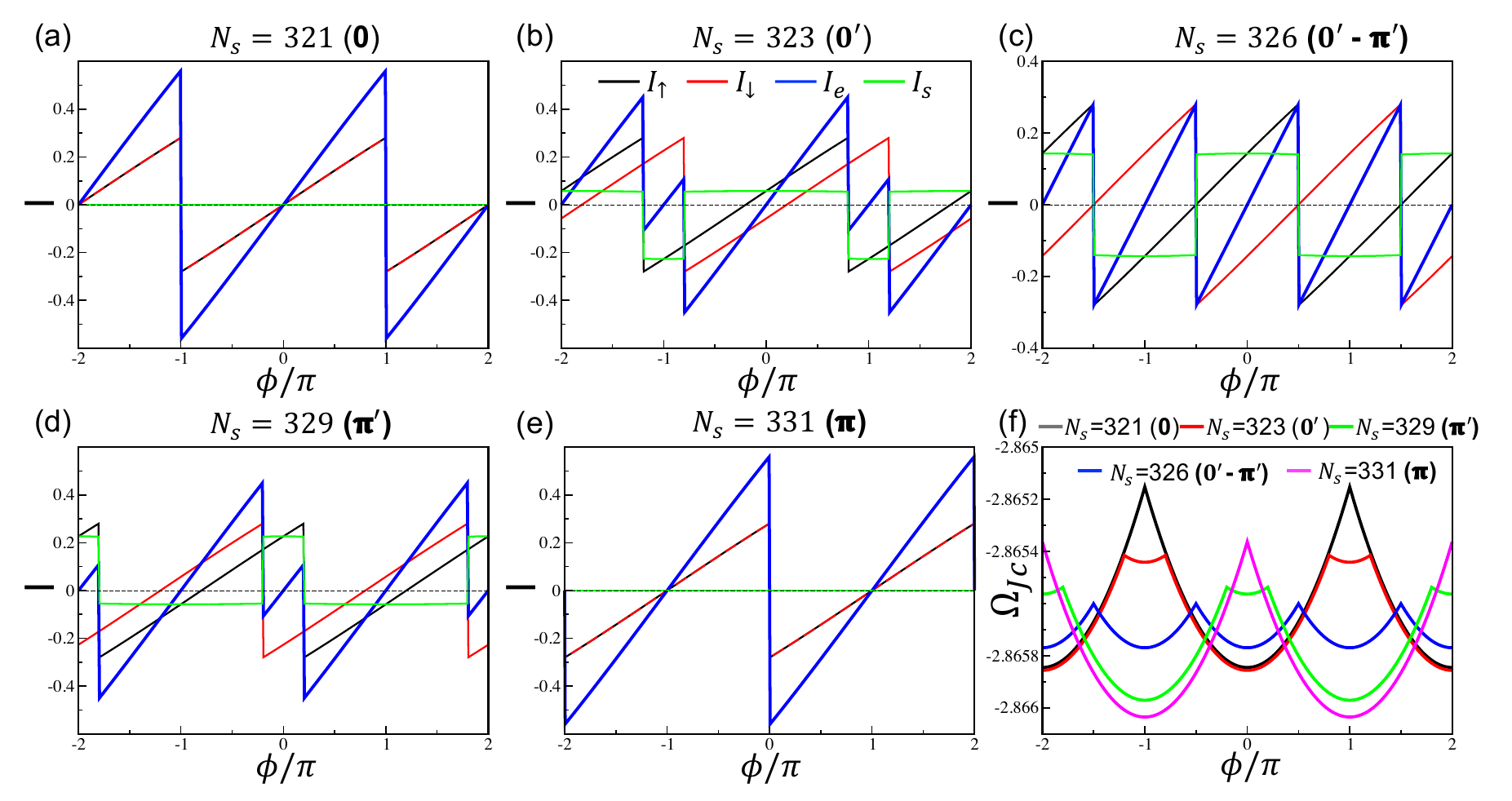}\caption{(a-e) Josephson current for different values of $N_{s}$. The parameters are $t_{1}=1$, $t_{2}=0$, $\alpha=\tan(\frac{\pi}{20})$, $\mu=0$, $V_1=2$ in the SC region. The KFF state has $Q=\frac{\pi}{10}$ and $\Delta_{\parallel}=0.169$. The normal region has $t_{N}=t_{L}=t_{R}=1$ as defined in Appendix.~\ref{appendixD}. (f) The total free energy at the corresponding values of $N_{s}$ in (a-e), showing various junction states labelled by $\mathbf{0}$, $\mathbf{0^{\prime}}$, $\mathbf{0^{\prime}}-\pmb{\pi}^{\prime}$, $\pmb{\pi}^{\prime}$ and $\pmb{\pi}$ determined by the distribution of the (local) minima at $\phi=0$ and $\pi$. The charge current $I_e(\phi)$ in (a-e) vanishes at both $\phi=0$ and $\pi$, the minima of the free energy in (f).
			\label{fig:0pi}}
	\end{center}
	\vskip-0.8cm
\end{figure*}

The dependence of the phase shift on the length $N_{s}$  of the SC region can be understood by performing a local gauge transformation that maps the KFF JJ onto a Kitaev JJ consisting of two spin-degenerate $p$-wave Kitaev chains subject to spin-dependent phase bias $\phi_{\sigma}=\phi+\sigma Q(N_{s}-1)$ as shown in Appendix.~\ref{appendixD}.
If we further assume the Josephson current for the Josephson chain consisting of two spin degenerate $p$-wave Kitaev chains with phase bias $\phi$ as $I_0(\phi)$ which is identical for the two spin species due to the spin degeneracy, we can then immediately get the Josephson current for each spin species as $I_{\sigma}(\phi)=I_{0}(\phi+\sigma Q(N_{s}-1))$, i.e., the Josephson current $I_{\sigma}(\phi)$ is shifted from the current of the transformed junction $I_0(\phi)$ by a phase $\sigma Q(N_{s}-1)$ (mod $2\pi$) so that the relative phase difference of the current between the two spin species is $\delta\phi=2\sigma Q(N_{s}-1)$ (mod $2\pi$).
Various Josephson junction states can be realized by tuning $\delta\phi$ from 0 to $2\pi$ through the length $N_s$.
This relation is verified numerically in Figs.~\ref{fig:0pi}(a-e) for $Q=\frac{\pi}{10}$ and $N_s\in [321,332]$. Any combination of $Q$ and $N_s$ can produce similar results as long as $\delta\phi$ covers the range $[0,2\pi]$.
We also calculate the total free energy of the system for 
different $N_{s}$ and indeed observe transitions between all these states controlled by $N_{s}$ as shown in Fig.~\ref{fig:0pi}(f). Specifically, for $N_{s}$=321 (331), there is only one global minimum located at $\phi=0\ (\pi)$ and the system is in $\mathbf{0}\ (\pmb{\pi})$ state. For $N_{s}$=323 (329), the free energy has a global minimum at $\phi=0\ (\pi)$ and a local minimum at $\phi=\pi\ (0)$ so that the system is in $\mathbf{0}^{\prime}$  ($\pmb{\pi}^{\prime}$) state. The transition between the two states is reached at $N_{s}=326$, where $\delta\phi=\pi$ and $\Omega_{Jc}(0)=\Omega_{Jc}(\pi)$.

Interestingly, in this
case with only nearest neighbor hopping $t_1$,  the 
charge current $I_{e}(\phi)$ acquires a period of $\pi$ in phase $\phi$ instead of $2\pi$ as in conventional Josephson current at the critical point ($N_s=326$), which can also be understood in the gauge transformed basis.
In this case with $N_s=326$, the phase bias becomes $\phi_\sigma=\phi+\frac{\sigma\pi}{2}$ and we thus have $I_{\sigma}(\phi)=I_{0}(\phi+\frac{\sigma\pi}{2})$, from which we can get $I_{\uparrow}(\phi+\pi)=I_{0}(\phi+\frac{3\pi}{2})=I_{0}(\phi-\frac{\pi}{2})=I_{\downarrow}(\phi)$.
Then we can derive a new relation for the total charge current $I_e$ as
\[
I_e(\phi+\pi)=I_{\uparrow}(\phi+\pi)+I_{\downarrow}(\phi+\pi)=I_{\downarrow}(\phi)+I_{\uparrow}(\phi)=I_e({\phi})
\] 
so that the charge current $I_e(\phi)$ acquires a period of $\pi$ instead of $2\pi$. The spin current $I_s(\phi)$ then acquires a minus sign when progressing $\pi$ phase, so that its period is still $2\pi$.
\[
I_s(\phi+\pi)=I_{\uparrow}(\phi+\pi)-I_{\downarrow}(\phi+\pi)=I_{\downarrow}(\phi)-I_{\uparrow}(\phi)=-I_s({\phi})
\]

The realization of diverse Josephson junction states by simply controlling the length of the superconductor is an intriguing property. 
It originates from the opposite nonzero momentum of Cooper pairs in each spin sector in the novel KFF state. The quantum interference of relative phase shifted pairing functions with opposite spin polarization leads to rich phases and physical phenomena.

\begin{table}[htb]
	\begin{tabular}{|c||c|}
		\hline
		State label & Distribution of minimums of the free energy  \\ \hline
		$\mathbf{0}$ & global minimum at $\phi=0$  \\ \hline
		$\mathbf{0^{\prime}}$ & global minimum at $\phi=0$ and local minimum at $\phi=\pi$ \\ \hline
		$\mathbf{0^{\prime}}-\pmb{\pi}^{\prime}$ & global minimum at both $\phi=0$ and $\phi=\pi$  \\ \hline
		$\pmb{\pi}^{\prime}$ & global minimum at $\phi=\pi$ and local minimum at $\phi=0$ \\ \hline
		$\pmb{\pi}$ & global minimum at $\phi=\pi$ \\ \hline
	\end{tabular}
	\caption{Definitions of the various Josephson junction states via the distributions of the minimums of the free energy.}
	\label{table1}
	\vskip-0.3cm
\end{table}

\section{Discussion}
\label{sec:dis}
We reported the theoretical discovery of a novel time-reversal invariant, finite momentum pairing Fulde-Ferrell state -- the KFF state. The concrete effective 1D model we used to realize the KFF state and its many unprecedented properties is intimately connected to the novel physics observed at the atomic line defect (ALD) in monolayer iron-based superconductor Fe(Te,Se), where zero-energy bound states emerge at both ends of  the ALD with no signatures of ${\cal T}$ symmetry breaking~\cite{chen_2020}. The missing atoms cause inversion symmetry breaking and induces Rashba SOC. It was shown that significant equal-spin triplet pairing can be induced by coherent quantum mechanical processes along such a Rashba ALD embedded in 2D unconventional superconductors ~\cite{yi_2021}.
This makes it plausible for materializing the effective 1D model with significant equal-spin triplet pairing to generate the KFF state.
More recently, evidence for finite momentum pair density wave order has been observed in monolayer Fe(Te,Se) along one-dimensional domain walls~\cite{liu}. The experimental evidence suggests time-reversal symmetry is preserved, which makes the KFF state a plausible candidate in addition to the Larkin-Ovchinnikov state.

The most remarkable of the KFF state is that, in the presence of broken inversion symmetry, it supports nonreciprocal spin supercurrent in both bulk superconductor and JJ.
In contrast to nonreciprocal charge transport in SC systems which requires breaking both inversion and ${\cal T}$ symmetry, nonreciprocal SC spin transport only requires breaking inversion symmetry.
This is because the spin current operator is invariant under time-reversal, such that systems with positive and negative spin current are unrelated
by the ${\cal T}$ operation. This is true regardless of whether the system respects the ${\cal T}$ symmetry or not, making it free of the constraint by the Onsager relation.
We thus propose a novel SC spin diode effect as a potential new frontier for using spins to make dissipationless electronic devices in SC spintronics.
While future work is clearly needed which is outside the scope of the current paper, we point out that the unique properties of the KFF state make it plausible for possible realizations in JJs consisting of a ferromagnetic barrier. The exchange field of the barrier favors spin-triplet pairing and has very little effect on the critical current of equal-spin triplet pairing such as in the KFF state, resulting in the slow decay of the critical current with increasing barrier length. Both effects have been demonstrated experimentally \cite{robinson_2010,khaire_2010}. In turn, detecting the nonreciprocal spin transport together with the slow decay of the critical current with the barrier length can serve as the smoking gun evidence for the KFF state.

\begin{acknowledgments}
We thank Kun Jiang for helpful discussions. YZ is supported in part by National Natural Science Foundation of China (NSFC) Grants No. 12004383, No. 12074276 and No 12274279. ZW is supported by the U.S. Department of Energy, Basic Energy Sciences, Grant No. DE FG02-99ER45747.
\end{acknowledgments}

\appendix

\section{More detailed results from the meanfield calculation}
\label{appendixA}

\begin{figure}[h]
	\begin{center}
		\fig{3.4in}{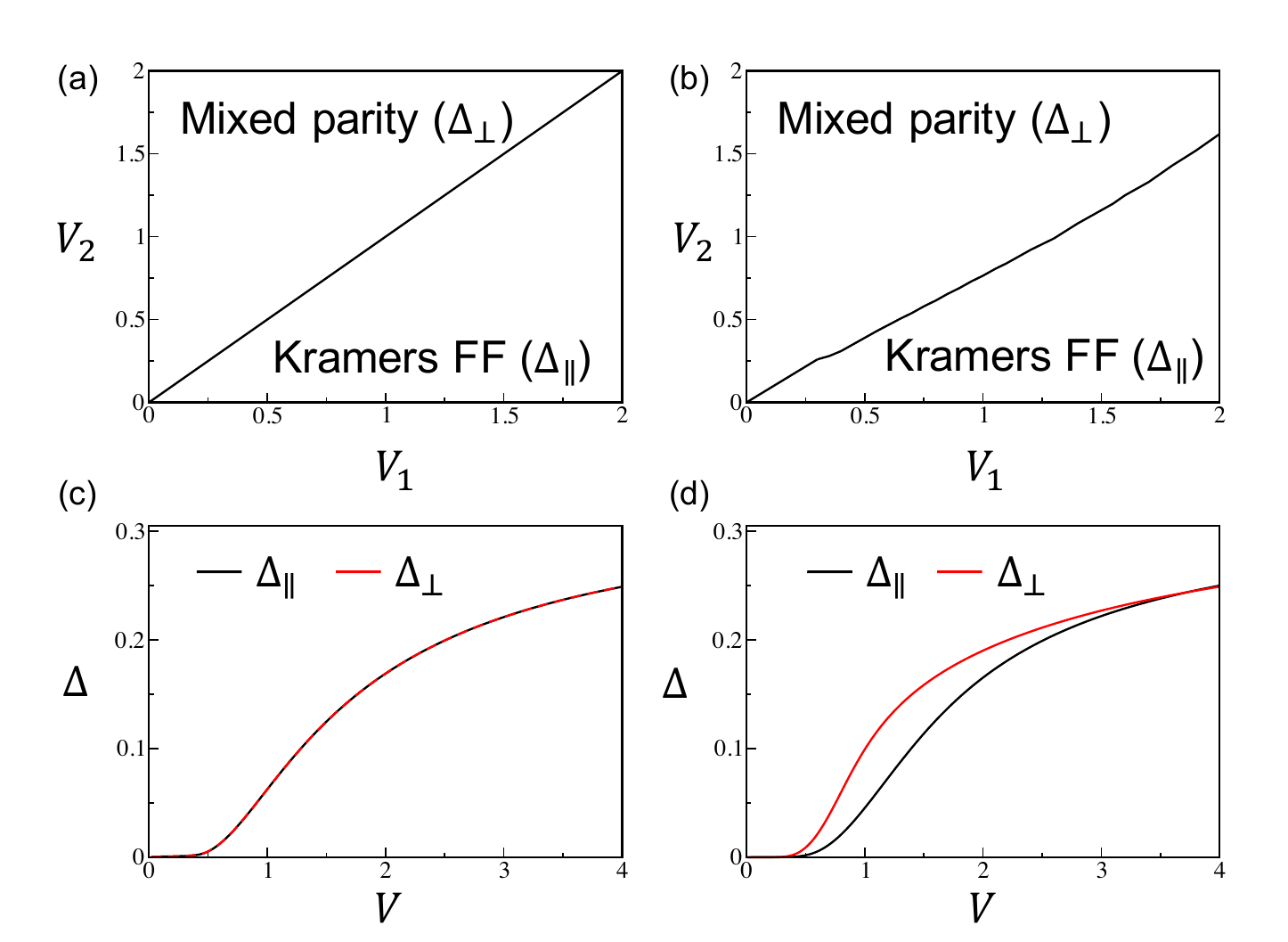}\caption{(a,b) Phase diagram determined from the meanfield calculations with two sets of parameters.  (c,d) The superconducting order parameter for both channels as the function of the attraction V showing the Cooper instability for the two sets of paramters. The parameters used are $t_{1}=1$, $t_{2}=0$, $\alpha=\tan(\frac{\pi}{20})$, $\mu=0$ for (a,c) and $t_{1}=1$, $t_{2}=-0.5$, $\alpha=0.4$, $\mu=-0.53$ for (b,d).\label{fig:pdsupp}}
	\end{center}
	\vskip-0.5cm
\end{figure}

We perform the calculation with two sets of parameters, which are shown in Fig.~\ref{fig:pdsupp}. As shown in Fig.~\ref{fig:pdsupp}, the ground states with finite pairing are either the KFF states with order parameters solely condensed in the equal-spin pairing channel or the mixture of $s$ and $p_{z}$ wave pairing states whose order parameters are solely condensed in the opposite-spin pairing channel and no mixed states with the coexistence of the order in both channels are found as the ground states except along the phase boundary of the two states where these two states are degenerate. This means that we can consider the two pairing channels separately, which further simplifies the meanfield Hamiltonian and is helpful for us in studying the properties of each state. 

\section{Derivation of the spin current for the KFF state}
\label{appendixB}

As shown in the main text, the meanfield Hamiltonian purely in the equal-spin pairing channel can be written as
\begin{equation}
	\begin{split}
	\hat{H}_{MF}-\mu\hat{N}&=\sum_{k\sigma}\left(\varepsilon_{k\sigma}-\mu\right)c_{k\sigma}^{\dagger}c_{k\sigma}
	\\
	&+V_{1}\Delta_{\parallel}\sum_{k\sigma}i\sin kc_{k+\frac{\sigma Q}{2},\sigma}^{\dagger}c_{-k+\frac{\sigma Q}{2},\sigma}^{\dagger}+h.c.+2N_{c}V_{1}\Delta_{\parallel}^{2}
	\end{split}
	\label{eq:mfsup}
\end{equation}
which can be further simplified in the Nambu basis $\psi_{k\sigma}^{\dagger}=\left(c_{k+\frac{\sigma Q}{2},\sigma}^{\dagger},c_{-k+\frac{\sigma Q}{2},\sigma}\right)$ as
\begin{equation}
	\hat{H}_{MF}-\mu\hat{N}=\frac{1}{2}\sum_{k\sigma}\psi_{k\sigma}^{\dagger}h_{k,Q,\sigma}\psi_{k\sigma}+2N_{c}V_{1}\Delta_{\parallel}^{2}-\mu N_{c} 
\end{equation}
with
\begin{equation}
	h_{k,Q,\sigma}=\left[\begin{array}{cc}
		\varepsilon_{k+\frac{\sigma Q}{2},\sigma}-\mu & 2iV_{1}\Delta_{\parallel}\sin k\\
		-2iV_{1}\Delta_{\parallel}\sin k & -\varepsilon_{-k+\frac{\sigma Q}{2},\sigma}+\mu
	\end{array}\right]
	\label{eq:hkff}
\end{equation}
In order to derive the expression for the spin current, let's consider the free energy density of the KFF state at finite temperature T, which is given as
\begin{equation}
	\begin{split}
	\Omega(\Delta_{\parallel},Q,T)&=-\frac{T}{N_{c}}\ln\text{Tr}\left[e^{-\frac{\hat{H}_{MF}-\mu\hat{N}}{T}}\right]
	\\
	&=-\frac{T}{2N_{c}}\sum_{k\sigma}\text{tr}\left[\ln(1+e^{-\frac{h_{k,Q,\sigma}}{T}})\right]+2V_{1}\Delta_{\parallel}^{2}-\mu
	\\
	&=\sum_{\sigma}\Omega_{\sigma}(\Delta_{\parallel},Q,T)-\mu
    \end{split}
\end{equation}
with
\begin{equation}
	\Omega_{\sigma}(\Delta_{\parallel},Q,T)=-\frac{T}{2N_{c}}\text{tr}\left[\ln(1+e^{-\frac{h_{k,Q,\sigma}}{T}})\right]+V_{1}\Delta_{\parallel}^{2}
\end{equation}
The current operator for each spin species in the system studied is defined as
\begin{equation}
	\hat{j}_{\sigma}=\frac{1}{N_{c}\hbar}\sum_{k}\partial_{k}\varepsilon_{k\sigma}c_{k\sigma}^{\dagger}c_{k\sigma}
\end{equation}
Next, from Eq.~\ref{eq:hkff} we have
\begin{equation}
	\partial_{Q}h_{k,Q,\sigma}=\left[\begin{array}{cc}
		\frac{\sigma}{2}\partial_{k}\varepsilon_{k+\frac{\sigma Q}{2},\sigma} & 0\\
		0 & -\frac{\sigma}{2}\partial_{k}\varepsilon_{-k+\frac{\sigma Q}{2},\sigma}
	\end{array}\right]
\end{equation}
and then we further have
\begin{equation}
	\begin{split}
		&\frac{1}{N_{c}\hbar}\sum_{k}\psi_{k\sigma}^{\dagger}\partial_{Q}h_{k,Q,\sigma}\psi_{k\sigma}=\frac{1}{N_{c}\hbar}\sum_{k}\Bigl[\frac{\sigma}{2}\partial_{k}\varepsilon_{k+\frac{\sigma Q}{2},\sigma}c_{k+\frac{\sigma Q}{2},\sigma}^{\dagger}c_{k+\frac{\sigma Q}{2},\sigma}
		\\
		&-\frac{\sigma}{2}\partial_{k}\varepsilon_{-k+\frac{\sigma Q}{2},\sigma}(1-c_{-k+\frac{\sigma Q}{2},\sigma}^{\dagger}c_{-k+\frac{\sigma Q}{2},\sigma})\Bigl]
		\\
		&=\frac{1}{N_{c}\hbar}\sum_{k}\sigma\partial_{k}\varepsilon_{k\sigma}c_{k\sigma}^{\dagger}c_{k\sigma}=\sigma\hat{j}_{\sigma}
	\end{split}
\end{equation}
which means
\begin{equation}
	\hat{j}_{\sigma}=\frac{\sigma}{N_{c}\hbar}\sum_{k}\psi_{k\sigma}^{\dagger}\partial_{Q}h_{k,Q,\sigma}\psi_{k\sigma}
\end{equation}
Then the current for each spin species can be calculated as
\begin{equation}
	\begin{split}
	j_{\sigma}(\Delta_{\parallel},Q,T)&=\frac{\text{Tr}\left[\hat{j}_{\sigma}e^{-\frac{\hat{H}_{MF}-\mu\hat{N}}{T}}\right]}{\text{Tr}\left[e^{-\frac{\hat{H}_{MF}-\mu\hat{N}}{T}}\right]}=\frac{\sigma}{N_{c}\hbar}\text{tr}\left[\partial_{Q}h_{k,Q,\sigma}f(h_{k,Q,\sigma})\right]
	\\
	&=\frac{2\sigma}{\hbar}\partial_{Q}\Omega_{\sigma}(\Delta_{\parallel},Q,T)
    \end{split}
\end{equation}
where $f(x)=(1+e^{x/T})^{-1}$ is the Fermi distribution function. 
Here, the operator $\hat{j}_{\sigma}$ is the density current, from which we can define the charge current operator as 
$\hat{j}_{e}=e(\hat{j}_{\uparrow}+\hat{j}_{\downarrow})$
and the operator for the spin current carrying spin polarization in $z$ direction as 
$\hat{j}_{s}=\frac{\hbar}{2}(\hat{j}_{\uparrow}-\hat{j}_{\downarrow})$. Therefore, we can conveniently define the unit for the charge and spin current as $\frac{e}{\hbar}$ and 1, so that both currents can be written in a similar format.
We finally arrive at the expression for the spin current with spin up polarization as
\begin{equation}
	j_{s}(Q)=\sum_{\sigma}\frac{\sigma\hbar}{2}j_{\sigma}=\partial_{Q}\Omega(\Delta_{\parallel},Q,T)
\end{equation}
This means the ground state with the optimized value of $Q=Q_{0}$ does not carry any net spin current as expected, since $\partial_{Q}\Omega(\Delta_{\parallel},Q_{0},T)=0$, and when a spin current $j_{s}$ is applied to the KFF state, the state with a different value of $Q$, satisfying $j_{s}(Q)=j_{s}$ is realized. Then the critical spin current for the positive and negative directions are determined by the maximum and minimum values of $j_{s}(Q)$ sustained by the superconducting state, which can be defined as $j_{s,c+}=\max_{Q}[j_{s}(Q)]$ and $j_{s,c-}=min_{Q}[j_{s}(Q)]$.
Moreover, due to the presence of ${\cal T}$ symmetry, $\Omega_{\uparrow}(\Delta_{\parallel},Q,T)=\Omega_{\downarrow}(\Delta_{\parallel},Q,T)$, which means $j_{\uparrow}(\Delta_{\parallel},Q,T)=-j_{\downarrow}(\Delta_{\parallel},Q,T)$, such that the charge current $j_{c}(Q)=\sum_{\sigma}j_{\sigma}$ always vanishes in the KFF state as expected and $j_{s}(Q)=\sum_{\sigma}\frac{\sigma}{2}j_{\sigma}=j_{\uparrow}$.

\section{Determining the optimized Q for KFF state}
\label{appendixC}
We first consider a simpler case with only nearest neighbor hopping, i.e. $t_2=0$. 
In this case, the dispersion of the noninteracting Hamiltonian becomes $\varepsilon_{k\sigma}=-2t_{1}\cos k-2\sigma\alpha\sin k=-2t_{\alpha}\cos(k-\sigma\theta_{\alpha})$ with $t_{\alpha}=\sqrt{t_{1}^{2}+\alpha^{2}}$ and $\theta_{\alpha}=\arctan(\frac{\alpha}{t_{1}})$, which determines the Fermi momentum as
\begin{equation}
	k_{f\sigma,\pm}=\sigma\theta_{\alpha}\pm\arccos(-\frac{\mu}{2t_{\alpha}})
\end{equation}
Then the quasiparticle energy Eq.~\ref{eq:ekpm} becomes 
\begin{equation}
	\begin{split}
	E_{k\sigma,\pm}&=-2\sigma t_{\alpha}\sin(\theta_{\alpha}-\frac{Q}{2})\sin k
	\\
	&\pm\sqrt{\left[-2t_{\alpha}\cos(\theta_{\alpha}-\frac{Q}{2})\cos k-\mu\right]^{2}+4V_{1}^{2}\Delta_{\parallel}^{2}\sin^{2}k}
	\end{split}
\end{equation}
and from Eq.~\ref{eq:omegak} and the ${\cal T}$ symmetry, we have the free energy density at zero temperature as
\begin{widetext}
\begin{equation}
	\Omega(\Delta_{\parallel},Q)=\frac{1}{2N_{c}}\sum_{k\sigma,n=\pm}E_{k\sigma,n}\Theta(-E_{k\sigma,n})+2V_{1}\Delta_{\parallel}^{2}-\mu
	=\frac{1}{N_{c}}\sum_{k,n=\pm}E_{k\uparrow,n}\Theta(-E_{k\uparrow,n})+2V_{1}\Delta_{\parallel}^{2}-\mu
\end{equation}
which leads to the expression for the spin current as
\begin{equation}
	j_{s}(Q)=\partial_{Q}\Omega(\Delta_{\parallel},Q)=\frac{1}{N_{c}}\sum_{k,n=\pm}\Theta(-E_{k\uparrow,n})\left\{ t_{\alpha}\cos(\theta_{\alpha}-\frac{Q}{2})\sin k+\frac{\left[t_{\alpha}\sin(\theta_{\alpha}-\frac{Q}{2})\cos k\right]\left[2t_{\alpha}\cos(\theta_{\alpha}-\frac{Q}{2})\cos k\right]}{E_{k\uparrow,n}+2t_{\alpha}\sin(\theta_{\alpha}-\frac{Q}{2})\sin k}\right\} 
\end{equation}
\end{widetext}
We can easily see that when $Q=Q_{0}=2\theta_{\alpha}$, $E_{k\sigma,\pm}=\pm\sqrt{\left[-2t_{\alpha}\cos k-\mu\right]^{2}+4V_{1}^{2}\Delta_{\parallel}^{2}\sin^{2}k}=\pm E_{k}$, then we have $j_{s}(Q_{0})=\frac{1}{N_{c}}\sum_{k,n=\pm}\Theta(-nE_{k})t_{\alpha}\sin k=\frac{2t_{\alpha}}{N_{c}}\sum_{k}\sin k=0$. Therefore, the ground state is characterized by the momentum $Q_{0}=2\theta_{\alpha}=\sigma(k_{f\sigma,+}+k_{f\sigma,-})$, which is also consistent with the Fermi surface of the noninteracting Hamiltonian, where the electrons around the two Fermi points $k_{f\sigma,\pm}$ within the same spin species pair together, giving rise to a finite Cooper pair momentum $Q_{0}$. We numerically calculate the zero temperature free energy as well as the current density as shown in Fig.~\ref{fig:je}(a, c) of the main text, which indeed confirms that the free energy density reaches the minimum at $Q=Q_{0}=2\theta_{\alpha}$ and the critical spin currents $j_{s,c\pm}$ are determined by the maximum and minimum values of the spin current density.

\section{The relation between the phase difference and $N_{s}$}
\label{appendixD}
To study the transport properties of the Josephson chain, we consider the system consisting of two superconductors with KFF order sandwiching a normal metal in between as shown in Fig.~\ref{fig:jj}(a) of the main text.
By setting the lattice constant to 1, this Josephson chain can be described by the tight-binding Hamiltonian $H_{Jc}(\phi)=H_{SL}+H_{N}+H_{SR}+H_{LN}+H_{RN}$, where
\begin{widetext}
\begin{equation}
	H_{SL}=-\sum_{i,j\in[1,N_{s}]}\sum_{\sigma}(t_{ij}-\delta_{ij}\mu)c_{i\sigma}^{\dagger}c_{j\sigma}
	+\sum_{i=1}^{N_{s}-1}\sum_{\sigma}\left(i\alpha\sigma c_{i\sigma}^{\dagger}c_{i+1\sigma}+\Delta_{\parallel}e^{i\sigma Qx_{i}+i\phi}c_{i\sigma}^{\dagger}c_{i+1\sigma}^{\dagger}\right)+h.c.
\end{equation}
\begin{equation}
	H_{SR}=-\sum_{i,j\in[N_{s}+N_{n}+1,2N_{s}+N_{n}]}\sum_{\sigma}(t_{ij}-\delta_{ij}\mu)c_{i\sigma}^{\dagger}c_{j\sigma}+\sum_{i=N_{s}+N_{n}+1}^{2N_{s}+N_{n}-1}\sum_{\sigma}\left(i\alpha\sigma c_{i\sigma}^{\dagger}c_{i+1\sigma}+\Delta_{\parallel}e^{i\sigma Q(x_{i}-N_{s}-N_{n})}c_{i\sigma}^{\dagger}c_{i+1\sigma}^{\dagger}\right)+h.c.
\end{equation}
\end{widetext}
describe the two SC regions on the left and right sides,
\begin{equation}
	H_{N}=-\sum_{i,j\in[N_{s}+1,N_{s}+N_{n}]}\sum_{\sigma}(t_{N,ij}-\delta_{ij}\mu)c_{i\sigma}^{\dagger}c_{j\sigma}
\end{equation}
describes the normal metal region in the middle and 
\begin{equation}
	H_{SNL}=-t_{L}\sum_{\sigma}c_{N_{s}\sigma}^{\dagger}c_{N_{s}+1\sigma}+h.c.
\end{equation}
\begin{equation}
	H_{SNR}=-t_{R}\sum_{\sigma}c_{N_{s}+N_{n}\sigma}^{\dagger}c_{N_{s}+N_{n}+1\sigma}+h.c.
\end{equation}
correspond to the coupling between the normal metal region and the left and right SC region, with $\phi$ the phase bias between the two SCs. Then the Josephson currents can be calculated by the formula
\begin{equation}
	I(\phi)=\frac{2e}{\hbar}\partial_{\phi}\sum_{n}f(\epsilon_{n})\epsilon_{n}(\phi)
\end{equation}
with $\epsilon_{n}$ the n-th eigenvalue for $H_{Jc}$ at the phase bias $\phi$ and $f(\epsilon)$ the Fermi distribution function. 

Next, we demonstrate the dependence of the relative phase difference on the length of the superconducting region $N_{s}$. Let us consider a simpler case with only nearest neighbor hopping on the superconducting region. We first perform a local gauge transformation
\begin{equation}
	\begin{cases}
		c_{i\sigma}^{\dagger}\rightarrow e^{-\frac{i\sigma}{2}Q(x_{i}-\frac{1}{2})}d_{i\sigma}^{\dagger} & \text{\text{for\ i\ensuremath{\in[1,N_{s}]}}}\\
		c_{i\sigma}^{\dagger}\rightarrow e^{-\frac{i\sigma}{2}Q(N_{s}-\frac{1}{2})}d_{i\sigma}^{\dagger} & \text{for\ i\ensuremath{\in[N_{s}+1,N_{s}+N_{n}]}}\\
		c_{i\sigma}^{\dagger}\rightarrow e^{-\frac{i\sigma}{2}Q(x_{i}-N_{n}-\frac{3}{2})}d_{i\sigma}^{\dagger} & \text{for\ i\ensuremath{\in[N_{s}+N_{n}+1,2N_{s}+N_{n}]}}
	\end{cases}
	\label{eq:gauge}
\end{equation}
then each term in Hamiltonian $H_{Jc}(\phi)$ becomes
\begin{widetext}
\begin{equation}
	H_{SL}=\sum_{i=1}^{N_{s}-1}\sum_{\sigma}\left[-t_{\alpha}e^{i\sigma(\frac{Q}{2}-\theta_{\alpha})}d_{i\sigma}^{\dagger}d_{i+1\sigma}+\Delta_{\parallel}e^{i\phi}d_{i\sigma}^{\dagger}d_{i+1\sigma}^{\dagger}\right]+h.c.-\mu\sum_{i=1}^{N_{s}}\sum_{\sigma}d_{i\sigma}^{\dagger}d_{i\sigma}
	\label{eq:sld}
\end{equation}
\begin{equation}
	H_{SR}=\sum_{i=N_{s}+N_{n}+1}^{2N_{s}+N_{n}-1}\sum_{\sigma}\left[-t_{\alpha}e^{i\sigma(\frac{Q}{2}-\theta_{\alpha})}d_{i\sigma}^{\dagger}d_{i+1\sigma}+\Delta_{\parallel}e^{-i\sigma Q(N_{s}-1)}d_{i\sigma}^{\dagger}d_{i+1\sigma}^{\dagger}\right]+h.c.-\mu\sum_{i=N_{s}+N_{n}+1}^{2N_{s}+N_{n}}\sum_{\sigma}d_{i\sigma}^{\dagger}d_{i\sigma}
	\label{eq:srd}
\end{equation}
\end{widetext}
\begin{equation}
	H_{N}=-\sum_{i,j\in[N_{s}+1,N_{s}+N_{n}]}\sum_{\sigma}(t_{N,ij}-\delta_{ij}\mu)d_{i\sigma}^{\dagger}d_{j\sigma}
\end{equation}
\begin{equation}
	H_{SNL}=-t_{L}\sum_{\sigma}d_{N_{s}\sigma}^{\dagger}d_{N_{s}+1\sigma}+h.c.
\end{equation}
\begin{equation}
	H_{SNR}=-t_{R}\sum_{\sigma}d_{N_{s}+N_{n}\sigma}^{\dagger}d_{N_{s}+N_{n}+1\sigma}+h.c.
\end{equation}
We can see that $H_{N}$, $H_{SNL}$ and $H_{SNR}$ are unchanged in the new basis, and if we further use the relation for the KFF state $Q=2\theta_{\alpha}$, the phase factors $e^{i\sigma(\frac{Q}{2}-\theta_{\alpha})}$ of the hopping $t_{\alpha}$ in Eq.~\ref{eq:sld}, \ref{eq:srd} disappear, and $H_{SL}$ and $H_{SR}$ then describe the spin degenerate p-wave Kitaev chains with superconducting phase $\phi$ and $-\sigma Q(N_{s}-1)$, which means $H_{Jc}(\phi)$ describes the Josephson chain consisting of two spin degenerate $p$-wave Kitaev chains with phase bias $\phi_{\sigma}=\phi+\sigma Q(N_{s}-1)$ for spin species $\sigma$.
If we further assume the Josephson current for the Josephson chain consisting of two spin degenerate $p$-wave Kitaev chains with phase bias $\phi$ as $I_0(\phi)$ which is identical for the two spin species due to the spin degeneracy, we can then immediately get the Josephson current for each spin species as $I_{\sigma}(\phi)=I_{0}(\phi+\sigma Q(N_{s}-1))$, i.e., the Josephson current $I_{\sigma}(\phi)$ is shifted from the current of the transformed junction $I_0(\phi)$ by a phase $\sigma Q(N_{s}-1)$ (mod $2\pi$) so that the relative phase difference of the current between the two spin species is $\delta\phi=2\sigma Q(N_{s}-1)$ (mod $2\pi$). We verify this relation numerically in Fig.~\ref{fig:0pi} of the main text where we take the parameters as $t_{1}=1$, $t_{2}=0$, $\alpha=\tan(\frac{\pi}{20})$, $\mu=0$, $V_1=2$, which leads to $Q=\frac{\pi}{10}$ and $\Delta_{\parallel}=0.169$.
For $N_{s}$ =321, 323, 326, 329 and 331 which leads to the phase difference $\delta\phi$ varying from 0 to $2\pi$, 
$I_{\sigma}$ is shifted by $0$, $\pm\frac{\pi}{5}$, $\pm\frac{\pi}{2}$, $\pm\frac{4\pi}{5}$ and $\pm\pi$, which is consistent with the results shown in Fig.~\ref{fig:0pi}(a-e) of the main text. 
Various Josephson junction states including $\mathbf{0}$, $\mathbf{0^{\prime}}$, $\mathbf{0^{\prime}}-\pmb{\pi^{\prime}}$, $\pmb{\pi^{\prime}}$ and $\pmb{\pi}$ junction states can be realized by tuning $\delta\phi$ from 0 to $2\pi$. The definition of these states is listed in Table.~\ref{table1} of the main text.

Moreover, if the second neighbor hopping $t_2$ is finite, after the gauge transformation of Eq.~\ref{eq:gauge}, Eq.~\ref{eq:sld},\ref{eq:srd} acquire extra term $-t_{2}e^{i\sigma Q}d_{i\sigma}^{\dagger}d_{i+2\sigma}$ as the 2nd neighbor hopping. Now, since the relation $Q=2\theta_{\alpha}$ no longer holds, neither this phase $e^{i\sigma Q}$ nor the phase $e^{i\sigma(\frac{Q}{2}-\theta_{\alpha})}$  of $t_{\alpha}$ in Eq.~\ref{eq:sld},\ref{eq:srd} can be gauged away, the transformed model no longer describes the Josephson chain consisting of two spin degenerate p-wave Kitaev chains, but rather two spin dependent p-wave Kitaev chains with complex hopping parameters that are time-reversal counterparts of each other, so that the time-reversal symmetry is not broken.

\section{$\Delta_{\perp}$ channel (mixture of $s$ and $p_{z}$ wave pairing state)}
\label{appendixE}
If we consider the meanfield Hamiltonian purely in the opposite-spin pairing channel, then the meanfield Hamiltonian becomes
\begin{equation}
	\begin{split}
	\hat{H}_{MF}-\mu\hat{N}&=\sum_{k\sigma}\left(\varepsilon_{k\sigma}-\mu\right)c_{k\sigma}^{\dagger}c_{k\sigma}
	\\
	&-2V_{2}\Delta_{\perp}\sum_{k}c_{k\uparrow}^{\dagger}c_{-k\downarrow}^{\dagger}\cos\left(k+\phi_{\perp}\right)
	\\
	&+h.c.+2N_{c}V_{2}\Delta_{\perp}^{2}
	\end{split}
\end{equation}
which can be further simplified in the Nambu basis $\psi_{k}^{\dagger}=\left(c_{k\uparrow}^{\dagger},c_{-k\downarrow}\right)$ as
\begin{equation}
	H_{MF}-\mu N=\sum_{k}\psi_{k}^{\dagger}h_{k}\psi_{k}+2N_{c}V_{2}\Delta_{\perp}^{2}-\mu N_{c}
\end{equation}
with
\begin{equation}
	h_{k}=\left[\begin{array}{cc}
		\varepsilon_{k\uparrow}-\mu & -2V_{2}\Delta_{\perp}\cos\left(k+\phi_{\perp}\right)\\
		-2V_{2}\Delta_{\perp}\cos\left(k+\phi_{\perp}\right) & -\varepsilon_{-k\downarrow}+\mu
	\end{array}\right]
\end{equation}
Diagonalizing $h_{k}$ , and considering the relation $\varepsilon_{k\uparrow}=\varepsilon_{-k\downarrow}=\varepsilon_{k}=-2t_{\alpha}\cos(k-\theta_{\alpha})-2t_{2}\cos(2k)$ owing to the ${\cal T}$ symmetry, we can get 
\begin{equation}
	E_{k,\pm}=\pm\sqrt{(\varepsilon_{k}-\mu)^{2}+4V_{2}^{2}\Delta_{\perp}^{2}\cos^{2}(k+\phi_{\perp})}=\pm E_{k}
	\label{eq:ekpmq0}
\end{equation}
Then the free energy density at zero temperature $\Omega(\Delta_{\perp},\phi_{\perp})$ can be calculated as
\begin{equation}
	\Omega(\Delta_{\perp},\phi_{\perp})=\frac{1}{N_{c}}\left\langle \hat{H}_{MF}-\mu\hat{N}\right\rangle =\frac{1}{N_{c}}\sum_{k,n=\pm}E_{k,n}\Theta(-E_{k,n})+2V_{2}\Delta_{\perp}^{2}-\mu
	\label{eq:omegakq0}
\end{equation}
with $\Theta(x)$ the Heaviside step function. Therefore, for a given value of $\phi_{\perp}$, the order parameter $\Delta_{\perp}$ can be determined self-consistently by minimizing $\Omega(\Delta_{\perp}$, $\phi_{\perp})$ with respect to $\Delta_{\perp}$, leading to the self-consistent equation
\begin{equation}
	\Delta_{\perp}=\frac{1}{N_{c}}\sum_{k} \cos\left(k+\phi_{\perp}\right) \left\langle c_{-k\downarrow}c_{k\uparrow}\right\rangle 
\end{equation}
The value of $\phi_{\perp}$ can be further determined by minimizing $\Omega(\Delta_{\perp}, \phi_{\perp})$ with respect to $\phi_{\perp}$, which is equivalent to have $\partial_{\phi_{\perp}}\Omega(\Delta_{\perp}, \phi_{\perp})=0$. From Eq.~\ref{eq:ekpmq0} and Eq.~\ref{eq:omegakq0}, we have
\begin{equation}  \partial_{\phi_{\perp}}\Omega(\Delta_{\perp},\phi_{\perp})=-\frac{1}{N_{c}}\sum_{k}\partial_{\phi_{\perp}}E_{k}=-\frac{V_{2}^{2}\Delta_{\perp}^{2}}{2\pi}\int_{-\pi}^{\pi}\frac{\sin(2k+2\phi_{\perp})}{E_{k}}dk
\end{equation}
If we further consider the case with $t_{2}=0$, then we have
\begin{widetext}
\begin{equation}
	\begin{split} \partial_{\phi_{\perp}}\Omega(\Delta_{\perp},\phi_{\perp})&=-\frac{V_{2}^{2}\Delta_{\perp}^{2}}{2\pi}\int_{-\pi}^{\pi}\frac{\sin(2k)}{\sqrt{\left[-2t_{\alpha}\cos(k-\phi_{\perp}-\theta_{\alpha})-\mu\right]^{2}+4V_{2}^{2}\Delta_{\perp}^{2}\cos^{2}k}}dk
		\\
		&=\frac{V_{2}^{2}\Delta_{\perp}^{2}}{2\pi}\int_{-\pi}^{\pi}\frac{\sin(2k)}{\sqrt{\left[-2t_{\alpha}\cos(k+\frac{\pi}{2}-\phi_{\perp}-\theta_{\alpha})-\mu\right]^{2}+4V_{2}^{2}\Delta_{\perp}^{2}\sin^{2}k}}dk
	\end{split}
\end{equation}
\end{widetext}
Apparently, when $\phi_{\perp}=\frac{n\pi}{2}-\theta_{\alpha}$ with integer n, the denominator of the integral is even in k while the numerator $\sin(2k)$ is odd in k, so that this integral vanishes, which means $\Omega(\Delta_{\perp}, \phi_{\perp})$ reaches extremum when $\phi_{\perp}=\frac{n\pi}{2}-\theta_{\alpha}$ and which one (odd n or even n) is the minimum depends on the details of the parameters. We note that when the free energy reaches a minimum at $\phi_{\perp}=\frac{\pi}{2}-\theta_{\alpha}$, the energy spectrum becomes $E_{k-(\phi_{\perp}-\frac{\pi}{2}),\pm}=\pm\sqrt{(-2t_{\alpha}\cos k-\mu)^{2}+4V_{2}^{2}\Delta_{\perp}^{2}\sin^{2}k}$ which is identical to the spectrum of the KFF state if $V_{1}=V_{2}$.

\bibliography{reference}

\end{document}